\documentclass[preprintnumbers,amsmath,amssymb,superscriptaddress,twocolumn,showpacs,etal,prl]{revtex4}
\pdfoutput=1
\usepackage{graphicx}
\usepackage{dcolumn}
\usepackage{bm}
\usepackage{times}
\begin{document}

\title{Non-Destructive Probing of Rabi Oscillations on the Cesium Clock Transition near the Standard Quantum Limit}

\author{P.~J.~Windpassinger}
\affiliation{QUANTOP, Niels Bohr Institute, University of
Copenhagen, Denmark}

\author{D. Oblak}
\affiliation{QUANTOP, Niels Bohr Institute, University of
Copenhagen, Denmark}

\author{P. G. Petrov}
\altaffiliation[Present address: ]{Department of Physics, Ben Gurion
University, Be'er-Sheva 84105, Israel}\affiliation{QUANTOP, Niels
Bohr Institute, University of Copenhagen, Denmark}

\author{M.~Kubasik}
\affiliation{ICFO, Institut de Ci\`{e}ncies Fot\`{o}niques, Barcelona,
Spain}\affiliation{QUANTOP, Niels Bohr Institute, University of
Copenhagen, Denmark}

\author{M.~Saffman}
\affiliation{Department of Physics, University of Wisconsin,
Madison, Wi, 53706 USA}

\author{C.~L.~Garrido Alzar}\altaffiliation[Present address:]{LCF, Institut d'Optique, 91127 Palaiseau, France}

\author{J. Appel}
\affiliation{QUANTOP, Niels Bohr Institute, University of
Copenhagen, Denmark}

\author{J. H. M\"{u}ller}
\affiliation{QUANTOP, Niels Bohr Institute, University of
Copenhagen, Denmark}

\author{N. Kj{\ae}rgaard}\email{niels.kjaergaard@nbi.dk}
\affiliation{QUANTOP, Niels Bohr Institute, University of
Copenhagen, Denmark}

\author{E. S. Polzik}
\affiliation{QUANTOP, Niels Bohr Institute, University of
Copenhagen, Denmark}

\date{\today}



\begin{abstract}
We report on non-destructive observation of Rabi oscillations on the
Cs clock transition. The internal atomic state evolution of a
dipole-trapped ensemble of cold atoms is inferred from the phase
shift of a probe laser beam as measured using a Mach-Zehnder
interferometer. We describe a single color as well as a two-color
probing scheme. Using the latter, measurements of the collective
pseudo-spin projection of atoms in a superposition of the clock
states are performed and the observed spin fluctuations are shown to
be close to the standard quantum limit.
\end{abstract}

\maketitle

The problem of a two-level atom interacting resonantly with a
coherent radiation field is of general importance to the field of
atomic and optical physics, and quantum optics \cite{Allen1987}.
Two-level quantum systems and the ability to manipulate them
coherently are essential to today's atomic clocks \cite{Vanier1989}
and tomorrow's quantum computers \cite{Zoller2005}. The coherence
life times in these systems can be measured via the decay of
population oscillations when subjected to an external resonant drive
field. Such Rabi oscillations are routinely observed in quantum dots
\cite{Stievater2001}, Josephson junction qubits \cite{Martinis2002},
nitrogen-vacancy centers in diamond \cite{Jelezko2004}, as well as
for trapped ions \cite{Leibfried2003} and atoms \cite{Matthews1998}.
In the case of trapped ions and atoms the internal quantum state
$|\!\!\uparrow \rangle$ or $|\!\!\downarrow \rangle$ is typically
detected via spontaneously scattered photons when probing the atoms
with laser light resonant with a transition from one of the states
to an auxiliary level. This method instantaneously destroys the
coherence between states $|\!\!\uparrow \rangle$ and
$|\!\!\downarrow \rangle$ interrupting the Rabi oscillations at the
time of probing. To sample the Rabi oscillations at given instances
of time requires at least as many re-preparations of the quantum
system \cite{Kuhr2005}. Rather than performing a completely
destructive measurement on an atomic sample, it is possible to gain
information about the quantum state in a more gentle fashion by
using off-resonant probe light and considering the ensemble of atoms
as a refractive medium. Such non-destructive probing could, e.g.,
prove advantageous to optical lattice clocks which are presently
limited by the time it takes to re-prepare atomic samples
\cite{Takamoto2005,Meiser2007}. Recently, a dispersive measurement
of Rabi oscillations on the cesium clock transition was demonstrated
\cite{Chaudhury2006}. In this work Rabi flopping between the Cs
clock states was observed by continuously probing the state
dependent birefringence of a cold atomic sample.  The authors state
that they were not able to reach the sensitivity sufficient for
observation of the atomic quantum spin noise which sets the ultimate
limit on sensitivity of such a measurement \cite{WINELAND1992}.
\begin{figure}[t!]
\includegraphics[width=0.85\columnwidth]{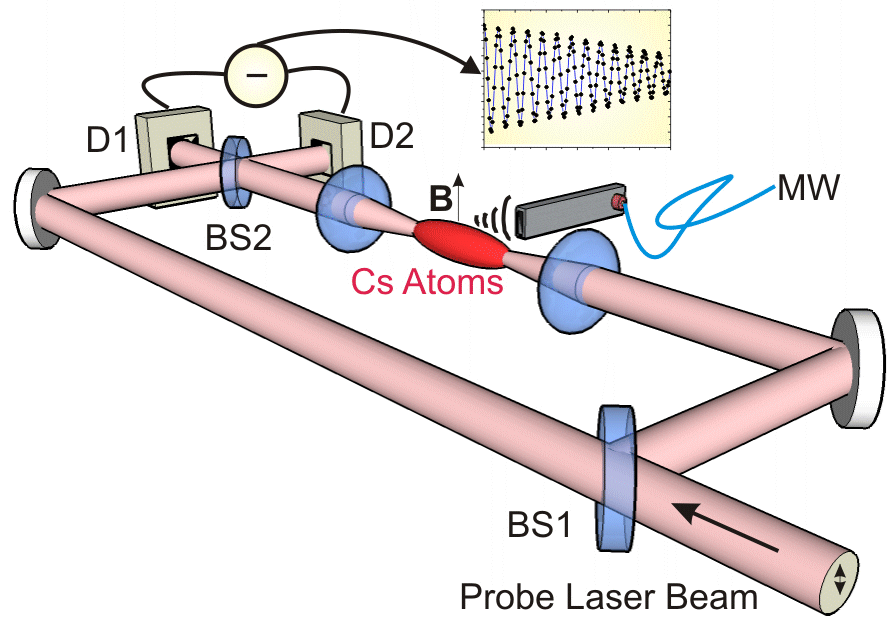}
\caption{(color online) Schematic drawing of the experimental setup.
A probe laser beam enters the Mach Zehnder via a port of the beam
splitter BS1. One part of the beam propagates through the Cs atoms
as they interact with the radiation field of the microwave source
MW. The other part of the beam propagates on a reference path with
no atoms. The two parts are recombined at the beams splitter BS2 and
the phase shift from the atoms is detected as the differential
signal of detectors D1 and D2 at the two output ports of
BS2.}\label{setup}
\end{figure}

In this Letter, we demonstrate a measurement of Rabi oscillations
and the pseudo-spin of the cesium clock transition which is fast,
nondestructive and allows to approach the standard quantum limit of
sensitivity. The state-dependent index of refraction of a cold
atomic ensemble is measured using a Mach-Zehnder (MZ) interferometer
(shown schematically in Fig.~1), which compares the phase
accumulated by a laser beam in a probe arm to that of a spatially
separated reference arm. The atoms are located in the probe arm and
modifies the optical path length according to their internal quantum
state. As compared to the polarization scheme \cite{Chaudhury2006}
the MZ interferometer provides a refractive index change per atom
which is larger since the reference beam is not phase shifted by the
atoms. We employ a pulsed probing scheme to take advantage of ac
photo detectors characterized by very low electronic noise compared
to typical dc detectors. This allows to reduce the number of probe
photons and thus to minimize the destructiveness of the atomic
measurement. In particular, our detection of light is shot noise
limited, which in turn allows for for the best detection of atomic
spin noise. Finally, rather than being released from a
magneto-optical trap (MOT), our atomic sample is confined by an
optical dipole trap resulting in a denser sample, available for an
extended period of time. As shown in \cite{Hammerer2004}, the
optical depth of the atomic ensemble is the key parameter for
optimum gain of information about the collective quantum state
versus its destruction.

Our experimental setup has been described in detail in
\cite{Petrov2007a}. In brief, Cs atoms are loaded from a MOT into an
optical dipole trap, where the atomic sample has a diameter of
$\sim\rm 60\,\mu m$ and contains up to $5\times10^5$ atoms,
achieving an optical depth around 25 as suggested by
\cite{Hammerer2004} .The trapped sample is located in one arm of our
MZ interferometer as shown in Fig.~1. The geometric path length
difference between the two arms of the interferometer is locked by a
piezo-actuated mirror using a weak auxiliary far off-resonance laser
beam counter-propagating the atomic probe beam. To minimize the
influence of the laser phase noise of the locking and the probe
beams, the interferometer is aligned to the so-called white light
position, where the two arms have equal length. After loading the
dipole trap we apply a small vertical bias field $\rm \sim\!1\,G$ to
define a quantization axis and optically pump the atoms to the
$(F=4,m_F=0)\equiv|\!\!\uparrow \rangle$ clock state using
vertically polarized laser light driving
$6S_{1/2}(F=4,m_F\neq0)\rightarrow6P_{3/2}(F'=4)$ $\pi$-transitions.
An additional repumping laser beam resonant with the
$6S_{1/2}(F=3)\rightarrow6P_{3/2}(F'=4)$ $\pi$-transitions ensures
that the atoms do not accumulate in the $F=3$ ground level. We end
up with up to 80\% of the atoms in $|\!\!\uparrow \rangle$ with an
efficiency likely to be limited by $m_F$-coherences created by our
narrow-bandwidth ($\sim250\,$kHz) pumping lasers
\cite{Tremblay1990}.

We address the clock states with 9.2~GHz microwave radiation
generated by a precision synthesizer. The orientation of the
linearly polarized magnetic field vector of the microwave field is
parallel to the guiding bias field of the atoms so that it only
drives magnetic dipole $\pi$-transitions. We have full control over
the power, frequency, relative phase, and duration of the microwave
field, so that we can produce any target state on the generalized
Bloch sphere of the effective two level system \cite{Allen1987}. As
discussed in \cite{Petrov2007a,Oblak2005} the phase shift of a light
beam with polarization state $q$ and wavelength $\lambda$
propagating a distance $l$ through an atomic gas with population
$N_{F,m_F}$ in each hyperfine sub-state $({F,m_F})$ confined to a
volume $V$ is
\begin{widetext}
\begin{eqnarray}
  \Delta \phi &= \phi_0    \sum_{F,m_F,F',m_F'} & N_{F,m_F} (2F'+1)(2F+1) {{\underbrace{\left(
                              \begin{array}{ccc}
                                F' & 1 & F \\
                                m'_F & q & -m_F \\
                              \end{array}
                            \right)}_{{\rm Wigner}\,3\!j\,{\rm symbol}}}^2
{\underbrace{\left\{
\begin{array}{ccc}
J & J' & 1 \\
F' & F & I \\
\end{array}
\right\} }_{{\rm Wigner}\,6\!j\,{\rm symbol}}}}^2   \frac{
    (\gamma/2)\Delta_{F,F'}}{\Delta_{F,F'}^2 + (\gamma/2)^2}, \label{phaseshiftformula}
\end{eqnarray}
\end{widetext}
where $\gamma$ is the line width of the transition, $\Delta_{F,F'}$
is the detuning of the incident light with respect to the
$L_J(F)\rightarrow L'_{J'}(F')$ transition, and
$\phi_0=3l\lambda^2(2J'+1)/4\pi V$; the Wigner $3\!j$ and $6\!j$
symbols are defined in \cite{sobelman}. The atom probing light is
derived from an external grating stabilized diode laser frequency
locked to the atomic line. The frequency of the probe laser is
$\Delta_{4,5} = 160$\,MHz blue detuned from the Cs D2 line $F=4 \to
F'=5$ transition. Since the phase-shift decreases with $1/\Delta$,
this probe light is only sensitive to the atomic population in the
$F=4$ state and the expression (\ref{phaseshiftformula}) reduces to
(assuming $m_F=0$ and light polarized along the quantization axis
$q=0$):
\begin{eqnarray}
  \Delta \phi = \frac{5}{36}\phi_0  N_{4,0}   \frac{
    (\gamma/2)\Delta_{F,F'}}{\Delta_{F,F'}^2 + (\gamma/2)^2}, \label{phaseshiftformulaspec}
\end{eqnarray}
which is the $|\!\!\uparrow \rangle$ clock state contribution to the
phase shift.

Atomic probing is performed with light pulses with a typical
duration of a microsecond produced with a standard acousto-optical
modulator. We record the pulses in a homodyne-like setup with a low
noise differential photo detector. The detector output is digitized
with a high bandwidth oscilloscope and the pulse areas extracted
using numerical integration. It is crucial for the minimal
destructiveness of the measurement that the observed phase
fluctuations of the probe light from the empty interferometer are
limited by the shot noise \cite{Petrov2007a}.
\begin{figure}[b!]
\includegraphics[width=0.79\columnwidth]{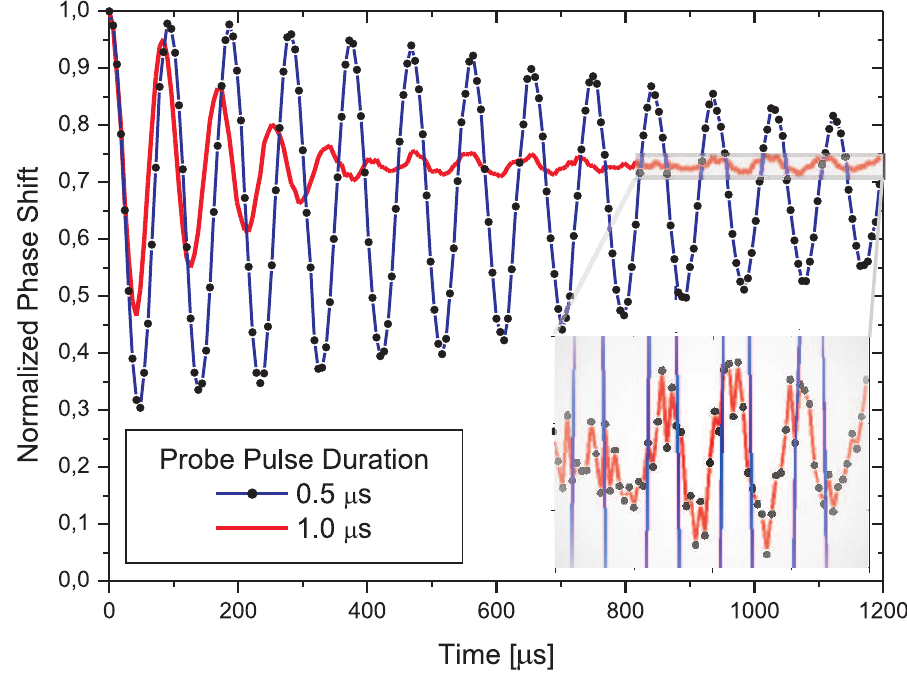}
\caption{(color online) Rabi oscillations (average of 10
experimental cycles) on the clock transition for an ensemble of Cs
atoms [optically pumped towards the $(F=4,m_F=0)$ state] as inferred
from the phase shift of probe laser light. The oscillations are
recorded using probe pulses with a repetition period of 6~$\mu$s and
duration of $0.5~\mu$s (blue line) and $1.0~\mu$s (red line),
respectively. In the latter case a pronounced damping of
oscillations is observed, however, with coherent dynamics at longer
times (see inset). In the former case the fringe contrast for the
normalized phase shift is limited to about 54\% as result of $\sim
30\%$ remaining atoms in $(F=4,m_F\neq 0)$ states.}\label{revival}
\end{figure}
\begin{figure*}[t!]
\includegraphics[width=0.88\textwidth]{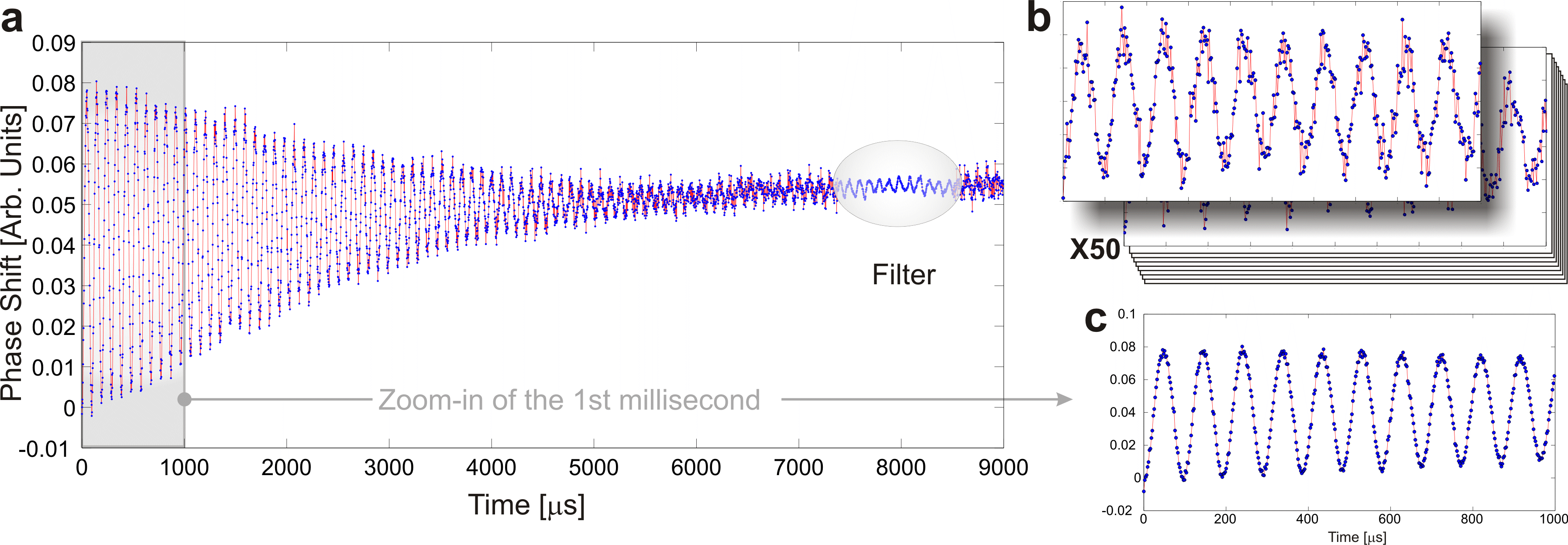}
\caption{(color online) Purified sample of atoms (a) Rabi cycling on
the clock transition recorded using 0.2~$\mu$s probe pulses with a
repetition period of 2.3~$\mu$s. Oscillations are still visible
after a time of 8~ms (a running average filter over 8~pulses has
been included at this point of time to enhance the oscillations)
corresponding to having applied $\sim3500$ probe pulses. The trace
is an average of 50 realizations of the experiment. (b) A single
experiment recording of Rabi oscillations during the first
millisecond. (c) A zoom-in on the first millisecond of the trace in
(a), i.e., an average of 50 traces as shown in (b). At time 0~ms the
phase shift is zero as result of all the atoms being transferred to
the $(F=3,m_F=0)$ clock state and the probe only being sensitive to
atoms in the $F=4$ states.} \label{zoom}
\end{figure*}
Figure \ref{revival} shows examples of microwave induced
Rabi-flopping between the clock states as observed with our
interferometer. The probe light beam interacting with the atoms has
a power of 140 nW distributed into $0.5~\mu$s and $1.0~\mu$s pulses
for the two traces shown, respectively (corresponding to
$3\times10^5$ and $6\times10^5$ photons/pulse). A marked difference
in oscillation frequency and decay between the two traces can be
observed as a result of the rather modest change in the probe pulse
duration by a factor of two. The decay of oscillation visibility is
caused by several processes. Even in the absence of the probe light
a decay will be present due to dephasing between atoms caused by
inhomogeneous differential light shifts from the dipole trapping
laser beam and small spatial variations in the microwave driving
field. The addition of probe laser light leads to a small amount of
spontaneous photon scattering which can be sub-divided into
inelastic Raman and elastic Rayleigh events \cite{Ozeri2005}. The
former leads to complete decoherence while the effect of the latter
depends on the internal state of the atoms in that atomic ensembles
at the poles of the Bloch sphere (i.e., all atoms in either the
$|\!\!\uparrow \rangle$ or the $|\!\!\downarrow \rangle$ state) are
not affected since elastic scattering takes an atom back to its
original internal state. Even with the probing being close to
non-destructive in the sense of little spontaneous photon
scattering, the dispersive interaction (responsible for the
observable phase shift of probe light) will act as to wash out Rabi
oscillations as a result of a frequency shift over the sample. An ac
Stark shift is imposed onto the atoms every time the probe light is
applied \cite{Featonby1998} changing the phase between clock states
superpositions as $|\!\!\downarrow \rangle+|\!\!\uparrow
\rangle\curvearrowright|\!\!\downarrow
\rangle+e^{i\phi}|\!\!\uparrow \rangle$. In our experiment the
average value of $\phi$ per photon is on the order of
$4\times10^{-7}$ rad which also characterizes the magnitude of the
phase spread over the atomic ensemble as a result of the Gaussian
probe beam profile. This inhomogeneous dephasing of atoms is
responsible for the dramatic increase in effective Rabi frequency
and envelope decay when doubling probe photon number in
Fig.~\ref{revival} \cite{tobe}. By only applying the probe pulses
every half oscillation period, in what corresponds to a spin echo
sequence, the dephasing can be minimized.

As mentioned, our optical pumping scheme is not able to transfer all
of the atoms to the $(F=4,m_F=0)$ clock state. In Fig.~\ref{revival}
this manifests itself as a fringe visibility limited to 54\%. To
purify the atomic polarization further we apply a microwave
$\pi$-pulse transferring these atoms to the
$(F=3,m_F=0)\equiv|\!\!\downarrow \rangle$ clock state and blow away
the remaining atoms in the $(F=4,m_F\neq0)$ states using resonant
light on the $6S_{1/2}(F=4)\rightarrow6P_{3/2}(F'=5)$ cycling
transition. In Fig.~\ref{zoom}(a) we show Rabi oscillation as
recorded for a such a purified ensemble initially in the
$(F=3,m_F=0)$ clock state. The atoms are probed with pulses
containing $\lesssim 10^5$~photons every $\rm 2.3\,\mu s$
corresponding to almost 50 times per Rabi cycle. Since this is far
beyond what can be resolved in the figure we include a zoom-in on
the first millisecond of evolution in Fig.~\ref{zoom}(b,c).
Figure~\ref{zoom}(b) shows a single experimental realization with
clear Rabi oscillations albeit with some noise. Of course, by
averaging over more experimental realizations [Fig.~\ref{zoom}(c)]
an improved signal to noise ratio is achieved. Clearly, the data
presented in Fig.~\ref{zoom}(b) demonstrates the capabilities of our
method in that the coherent evolution of an ensemble can be followed
``non-destructively" in ``real time''.
\begin{figure}[tb!]
\includegraphics[width=0.81\columnwidth]{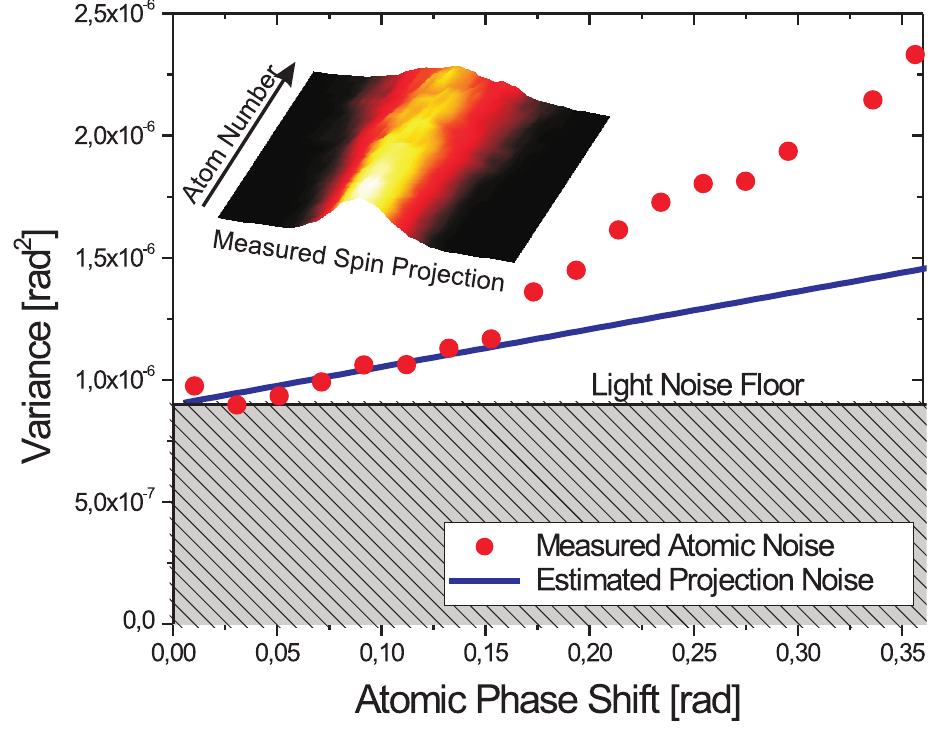}
\caption{(color online) The fluctuations in the interferometer phase
for ensembles in an equal superposition of the clock states as a
function of the interferometer phase shift with all atoms at a pole
of the Bloch sphere (this phase shift is proportional to the atom
number). The quantum projection noise estimated from our
experimental parameters is shown as a line. The inset shows the
distribution of the measured pseudo-spin projection, its width
increasing with atom number.} \label{noise}
\end{figure}

As is well known, a two-level quantum system is formally equivalent
to a spin $1/2$ particle so that our phase shift measurement of the
atomic ensemble can be interpreted as a collective pseudo-spin
projection observation. Experimentally, we can prepare an ensemble
of atoms in an equal coherent superposition of $|\!\!\uparrow
\rangle$ and $|\!\!\downarrow \rangle$ by applying a microwave
$\pi/2$ pulse to our atomic sample after the purification stage. It
is possible to achieve a zero phase shift for such a coherent spin
state (CSS) in the balanced interferometer by employing a two-color
probing scheme. To determine the population in $|\!\!\downarrow
\rangle$, we use an additional probe laser at $\Delta_{3,2}=
-135$\,MHz, red detuned from the $F=3 \to F=2$ transition and
coupled to the interferometer via the same optical fiber as the
$F=4\rightarrow F'=5$ probing laser. The detunings are arranged such
that with equal powers of the two probes, we obtain a zero mean
phase shift for equal populations in both levels, irrespectively of
the number of atoms in the ensemble. Using both probe beams (colors)
at the same time thus gives information on the population number
difference of the two hyperfine states. It is known
\cite{Hammerer2004,Oblak2005} that when the probe power is increased
the pseudo-spin measurements become limited by the projection noise
of atoms (the quantum fluctuations of the CSS
\cite{Santarelli1999}). In order to reach this regime we employ a
single pulse with $3.6\times10^7$ photons roughly corresponding to
the total photon number used during the entire observation cycle of
Fig.\ref{zoom}(b). Figure \ref{noise} shows the observed atomic
pseudo-spin fluctuations as a function of the atom number along with
an estimate of the atomic quantum projection noise. The data
presented results from the analysis of $\sim 34000$ repetitions of
the experimental cycle. As can be seen, for low atom numbers our
measurements of the collective pseudo-spin projection show the
expected quantum noise level and characteristic linear scaling of
atomic noise (variance) with atom number. For higher atom numbers,
however, non-quantum noise sources contribute by a non-negligible
part. We attribute this to the relative amplitude and phase
fluctuations of the two independent probes. Reduction of this
classical (quadratic) noise component will take us to the regime
where the atomic projection noise becomes comparable to the shot
noise of the probe light, which would be of considerable interest
since our dispersive light-atom interaction has a quantum
non-demolition (QND) character \cite{Kuzmich1998}. A measurement of
the collective pseudo-spin projection could hence be used to predict
the outcome of a subsequent measurement beyond the standard quantum
limit. Such predictive power squeezes the variance of the projection
conditionally. When used in a feed-back scheme the information
gained in the first measurement can be used to construct
unconditionally squeezed target states. Such engineered
non-classical collective states open up the possibility to surpass
the projection noise limit for atomic clocks operated using a Ramsey
sequence as discussed in \cite{Oblak2005} for our experimental
configuration.

In conclusion, we have measured the state dependent refractive index
of a trapped ensemble of Cs atoms and shown the ability to follow
coherent processes such as Rabi oscillations in ``real time". Using
a shot noise limited two-color probing scheme we have measured the
noise of the pseudo-spin projection of atoms in a superposition
between the clock states. The recorded atomic pseudo-spin noise is
near the standard quantum limit --– the projection noise, indicating
our ability to perform a QND measurement on cold dipole trapped
atoms.

This work was funded by the Danish National Research Foundation, as
well as the EU grants QAP, COVAQUIAL and EMALI. N.K. acknowledges
the support of the Danish National Research Council through a Steno
Fellowship and helpful advise from Stefan Kuhr.

\end{document}